# Effect of a role model-based show on the STEM aspirations of primary school students

Sadler, W.J.[a]*, Bartlett, S.[b]. and Dinmore, N.[c]
[a]*School of Physics & Astronomy, Cardiff University, Cardiff, UK*
sadlerwj@cardiff.ac.ukc
[b] School of Social Sciences, Cardiff University
[c] *CERN, Communications Officer – work done during undergraduate placement at School of Physics & Astronomy, Cardiff University*

**ABSTRACT**

A STEM outreach show called 'Who wants to be a Superhero' went into primary schools across England and Wales, delivered by professional science communicators. The show was structured around four female STEM role models and two male STEM role models as part of a wider narrative about scientists and engineers being real-life superheroes. The show was found to increase both boys' and girls' aspirations for a STEM career, with a bigger effect seen in the girls. The intervention also increased positive attitudes, and decreased negative attitudes, towards scientists measured through word association given by the audiences pre- and post-show. There was a larger reduction in negative stereotypes from the girls after the show, especially in the 'nerdy' and 'crazy' stereotypes. The show saw an increase in both boys' and girls' perceptions of the 'helpful' nature of science and engineering. It was also found that some students may not want to do a STEM job because of the perceived dangerous nature of the job.

**INTRODUCTION**

Many studies have been carried out over the past century regarding the uptake of STEM careers. Whilst demand for jobs in STEM has increased over time, the number of employment opportunities still outnumbers those pursuing careers in the field, especially in physical sciences and engineering (*Big Ideas: The Future of Engineering in Schools*, n.d.). Women are significantly under-represented in STEM fields, making up just 23% of those in core STEM occupations in the UK and 24% working in core STEM industries. In engineering, the figure is just 12% (Zenker, 2023). To combat these gender differences, many interventions have been attempted, yet these numbers have not changed significantly in the last 20 years (Clark Blickenstaff*, 2005)

**Role models and the connection to career choice**

In IMECHE's 2016 study (*Big Ideas: The Future of Engineering in Schools*, n.d.), the highest perceived reason for girls' career decisions and the lack of females in engineering was a 'lack of female role models'. Blickenstaff's (2005) literature review categorises explanations for the 'leaky pipeline' in science careers. The absence of female scientists/engineers as role models is listed as one of the most cited reasons for the leaky pipeline. Another study found the influence of role models has a small but significant influence on career choice of 18-25 year old women, above the influence of self-efficacy (Quimby & De Santis, 2006). It was found that a scientist was more likely to be a role model if the students could connect on things like personality and shared interests (Buck et al., 2008).

**Attributes of role models**

Stereotypes about scientists have been a 'hard-to-shift' issue in STEM recruitment since research into this area began. The 'Draw a scientist' (or DAS) test (Chambers, 1983) has been used for decades, and despite a small shift showing more female representation, the core characteristics used have changed little (Miller et al., 2018). Narrow stereotypes of scientists as white, male and highly intelligent still prevail, and even when positive attributes are recognised (e.g. around altruism), they can be described in a way that 'others' the scientist, showing that they are not seen as 'someone like me' (Shimwell et al., 2021).
In a study on a teacher intervention with primary school students, Shimwell et al (2021) found that by pro-actively using the attributes of a different STEM role model each week they could change some of the negative associations held about whether students felt they could become a scientist. This could be achieved indirectly without the scientist role model visiting the school. Many of these positive changes were retained a year later, but they found the idea of superior intelligence was a more difficult change to retain long-term.
In the evaluation of the WISE (Women in Science and Engineering) intervention called 'People like me'; student characteristics were used to connect girls with case studies of those who have jobs in STEM (Herman et al., 2023). Using this approach was shown to be effective at changing attitudes to STEM, with 57% of those involved saying they were 'more interested in studying STEM at school' after the event, and those not interested at all dropping from 10% to 4%.

**Purpose of Study**

From the literature we can see that stereotypes about STEM can be hard to shift but that role models do play an important part of the puzzle in terms of career aspirations. In this case study we propose that role models can be used effectively in an indirect way (through their inclusion in a science show) to change perceptions of, and self-identity with STEM.

**AIMS**

This case study aimed to investigate whether using role models in a science show context improved attitudes to science careers for students aged nine to eleven. The two main objectives of the research were:

1. To explore how watching a live show featuring videos of role models might change the stereotypical views about scientists and engineers among primary school students
2. To explore whether this indirect role-model intervention could change student aspirations to STEM careers
3. To exploring the relationship between the gender of the participants and the gender of the preferred role models/job aspirations

## METHODOLOGY

### Description of the science show intervention

'Who wants to be a Superhero?' is a 60-minute interactive STEM intervention designed for primary school students aged 9-11years old. It was conceived and developed by the social enterprise, Science Made Simple, with support from Women in Science and Engineering (WISE) in Wales and the Welsh Government National Science Academy scheme. The show uses an opening question asking students who their favourite superhero is, and what powers they have. A presenter then connects these answers to our 6 real-life "science heroes" and shows a short video of them talking about their job and their hobbies. Each section of the show then uses active-learning techniques and demonstrations to connect the job of the role model to familiar topics from the school curriculum. The science heroes, their job and 'superpower', along with the order they are introduced in the show, are summarised in Table 1.

| Role model pseudonym (gender)* | Job (simple summary of job given on questionnaire) | Superpower (related to narrative of the show) | Relevant comments about themselves, or reasons given for choosing their job |
|---|---|---|---|
| Harriet (f) | Astronomer (studies stars) | X-Ray vision | Chose because she didn't know what to do and teacher gave her a book about space which got her curious |
| Lola (f) | Civil engineer (builds bridges) | Strength | Enjoyed maths and physics at school. Talks about roads and bridges (in hard hat) |
| Hannah (f) | Medical engineer (keeps people safe) | Saving lives | Chose subject because good at art and design at school and want to make things that keep people safe |
| Libby (f) | Environmental scientist (helps the environment) | Saving the planet | Talks about a bus running on waste and sewage (the poo bus!) Makes a link to preventing animals becoming endangered |
| Evan (m) | Electronic engineer (builds lasers) | Lasers | Wanted to do something practical, plays with circuits. Systems fly on drones and go into space |
| Rory (m) | Aeronautical engineer (improves aeroplanes) | Flight | Did work experience with engineers, grew up with pilots in the family – learning to fly. Passion for planes. Doing what I love as a job is great |

**Pseudonyms are used to protect identity of individual scientists*

*Table 1: Role models used in the show and samples of the comments they include in their videos for context*

### Data Collection

A mixed methods approach was taken involving both quantitative and qualitative data. Written questionnaires were issued to students before and after the show. The pre-show questionnaire asked 3 questions, and the post-show questionnaire asked these again plus some additional questions about the role models in the show. The questionnaire responses were compiled using SPSS software and MS excel for quantitative analysis.

A few weeks after the show, four mixed-sex focus groups of between 5 and 8 students each from two different schools were held. The person conducting the interviews was the lead author (Wendy Sadler) and not the person who had presented the show as it was felt that could have biased the responses given due to social desirability bias (Bispo Júnior, 2022).

They were designed to be semi-structured, and the audio of these interviews was recorded, transcribed, and coded using NVivo software, with specific focus on attitudes towards scientists; responses to the role models; and awareness of science and gender stereotypes.

## RESULTS AND DISCUSSION

The total number of respondents to the questionnaire was 363 which was made up of 191 boys; 157 girls; and 15 of unspecified gender. The ages of all respondents were between nine and eleven. With questions connected to gender we did not include the 'unspecified' category in our results due to too few responses.

The focus groups were conducted in two randomly selected schools with a total of 25 students. The focus groups lasted between twenty to forty minutes each and these resulted in a total recording time of 102 minutes.

### STEM Job Aspirations pre- and post-intervention

Participants were asked both before and after the show if they 'would like a job in science or engineering when you are older?' and could respond with 'yes', 'no', or 'maybe'. Responses are summarised in Fig.1.

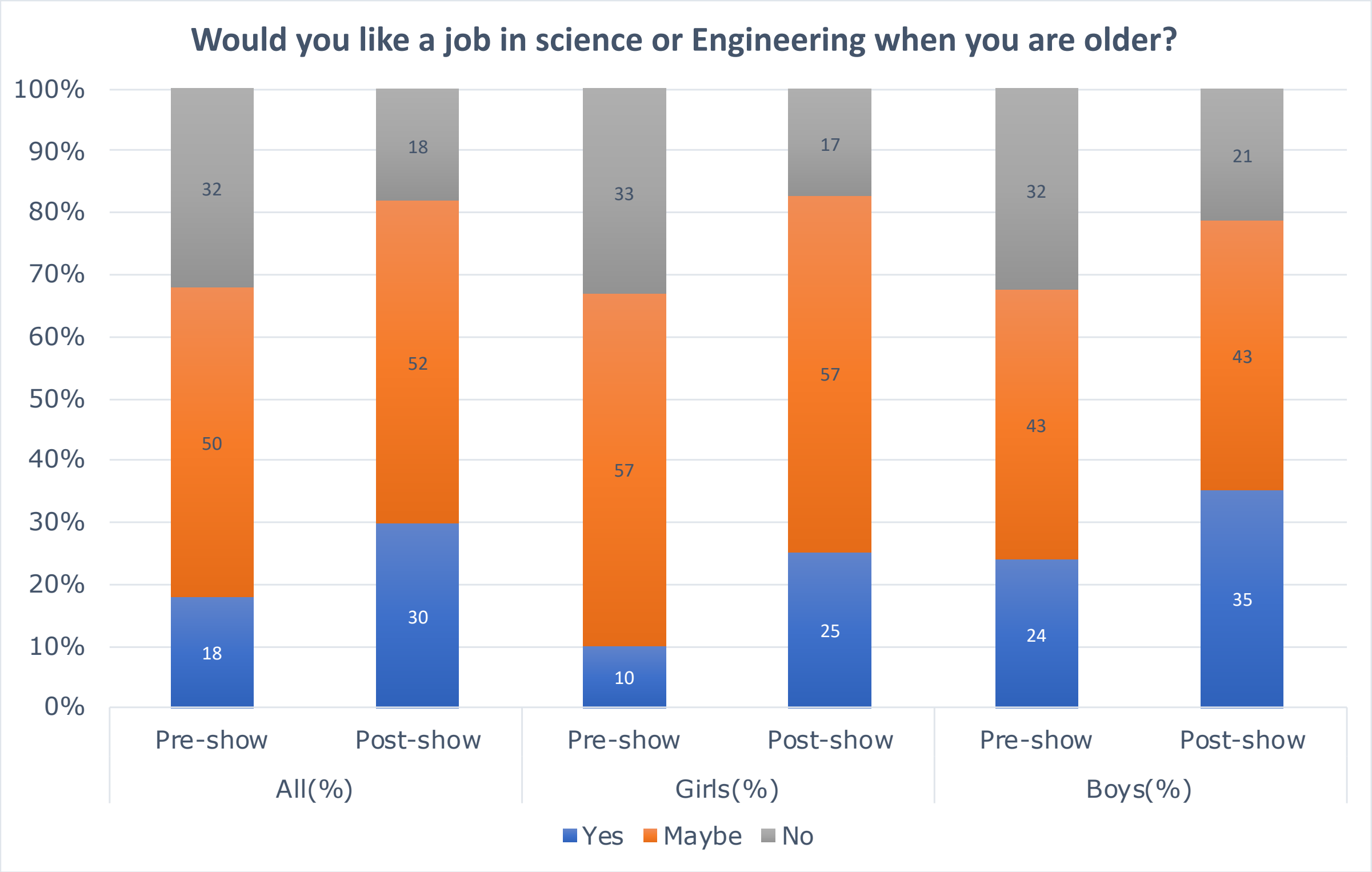

*Fig 1: Results before and after show for question; "Would you like a job using science or engineering when you are older?"*

Approximately 18% of all participants indicated they would like a career in science and engineering before the show, this is comparable to the benchmark data from ASPIRES which puts the figure of those wanting a STEM career at age 10 at about 17% (DeWitt & Archer, 2015). It is apparent that before the show, the proportion of both male and female participants that said 'no' were comparable, however more males answered a definite 'yes' than females, with more females saying 'maybe'.

Overall the show had an positive effect on students' STEM career aspirations. Before the show, only 18% of the 364 students said 'Yes' to the question 'Would you like a job using science or engineering when you are older?', and this rose by 12 percentage points to 30% after the show. Those who said they did not want a job in STEM also dropped from 32% (n=117) to 19% (n=67); a drop of 13 percentage points.

Looking in more detail at how the responses changed after the show, 14% moved from saying 'no' to 'yes' and 49% moved from saying 'no' to 'maybe'. Of those who said 'maybe' in pre-test, 25% changed to say 'yes' in the post-test, and 12% changed to say 'no' (Figure 2).

When looking at results according to gender, of the 61 males who responded 'no' in the pre-test, 16% said 'yes' and 46% said 'maybe' in the post-test'. Alternatively, for female participants, of those who responded 'no' in the pre-test, 10% said 'yes' in the post-test and 53% said 'maybe'. Of those who said 'maybe', 24% said 'yes' in the post-test. These results are summarised in Table 3.

| Pre- | Post- | All | Males | Females |
|---|---|---|---|---|
| Stayed the same across occasion | | 202 (57.4%) | 104 (55.3%) | 91 (59.1%) |
| No | No | 43 (37.4%) | 23 (37.7%) | 19 (37.3%) |
| | Maybe | 56 (48.7%) | 28 (45.9%) | 27 (52.9%) |
| | Yes | 16 (13.9%) | 10 (16.4%) | 5 (9.8%) |
| Maybe | Maybe | 110 (63.2%) | 46 (56.8%) | 59 (67.8%) |
| | No | 21 (12.1%) | 14 (17.3%) | 7 (8.0%) |
| | Yes | 43 (24.7%) | 21 (25.9%) | 21 (24.1%) |
| Yes | Yes | 49 (77.8%) | 35 (76.1%) | 13 (81.3%) |
| | Maybe | 11 (31.7%) | 8 (17.4%) | 3 (18.8%) |
| | No | 3 (4.8%) | 3 (6.5%) | 0 (0.0%) |

*Table 2: Showing movement on opinion on jobs in science before and after the show*

Slightly more female participants maintained their initial view after experiencing the show, approximately 60% compared to 55% of males. As is apparent from Figure 2, the most common change among participants' views was from a 'no' to a 'maybe'. The greatest change was seen among females with nearly 53% (n=27) making this change. Across both genders the group that least frequently showed a change of mind were those who said 'yes' before the show (and stuck to 'yes' after the show).

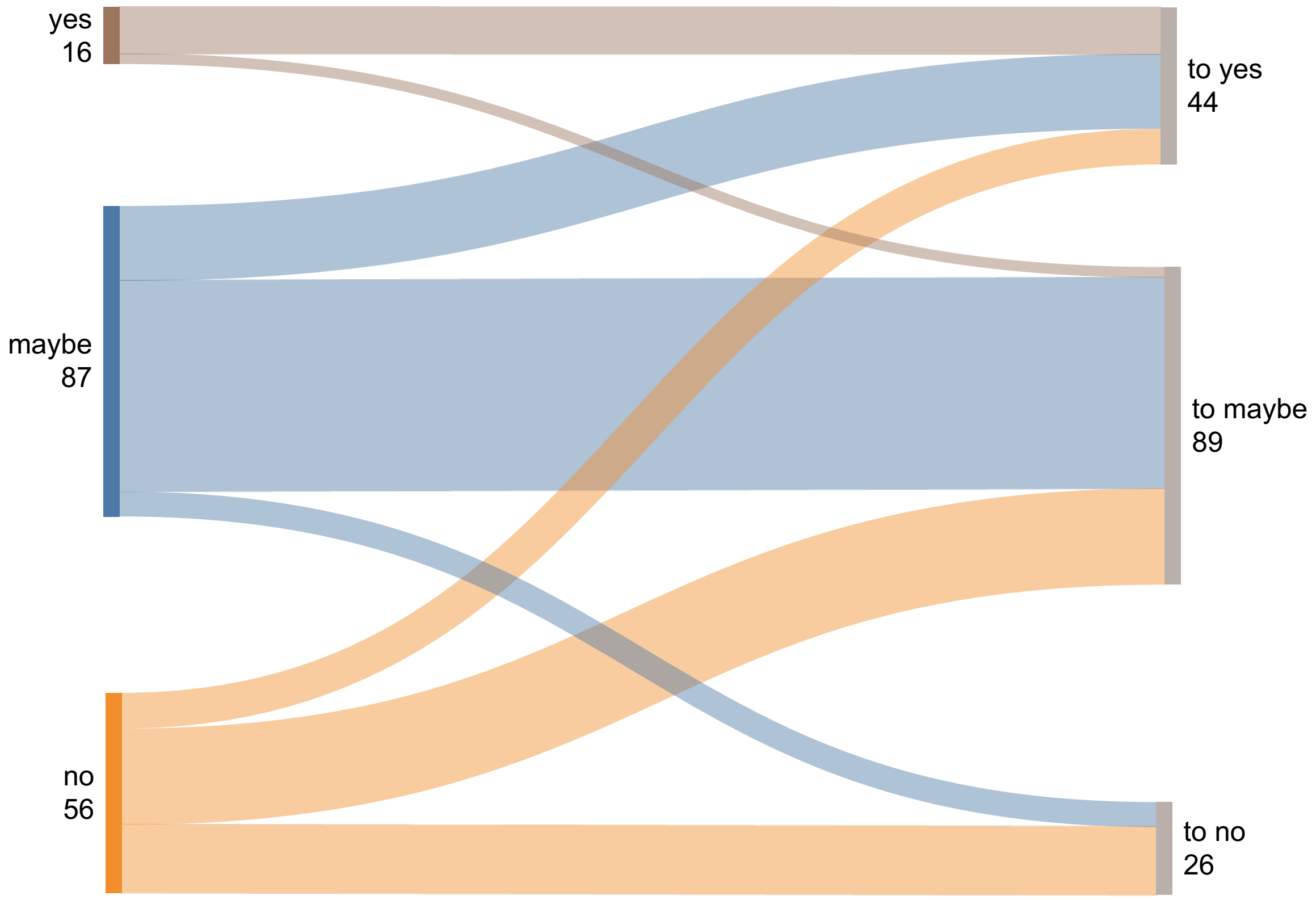

*Fig 2a: Showing girls' changes of opinion on wanting a job in science and engineering before and after the show*

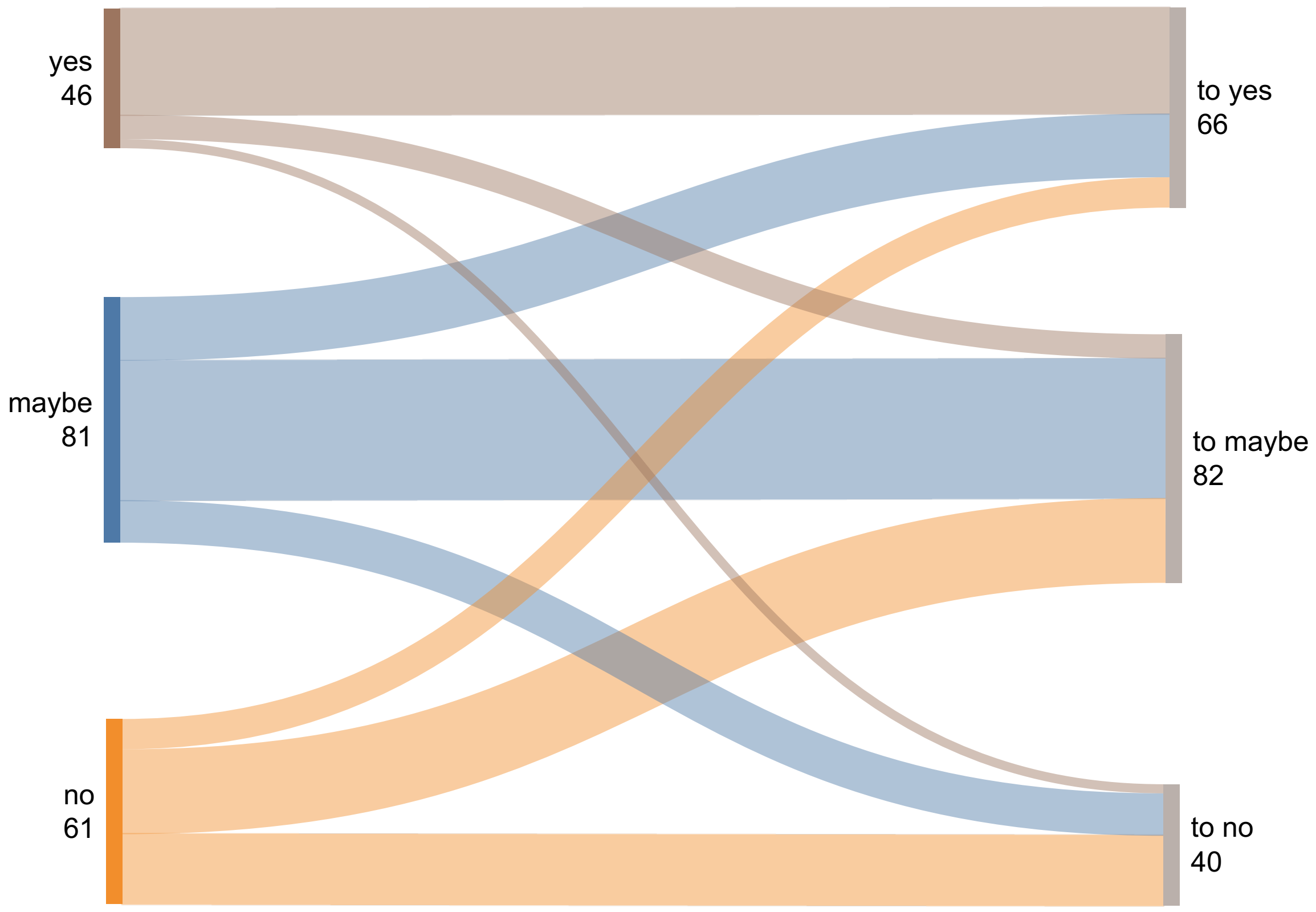

*Fig 2b: Showing boys' changes of opinion on wanting a job in science and engineering before and after the show*

**Reasons given for wanting/not wanting a job in science or engineering**
The results from the career aspirations questions were explored further through the follow-up open question; 'Why do you think that [about having job in science]?' Answers were categorised into altruistic motives, enjoyment, aptitude, taking after family, danger and miscellaneous/other. The main reason for wanting to do a science job was the perceived enjoyment. It was also the main reason for not wanting to do a science job (because the respondent had a preference or enjoyment for another subject or viewed science as boring). This supports previous research that suggests enjoyment of a subject is correlated with wanting a job in that area (Caleon & Subramaniam, 2008)

These justifications were also explored in the focus groups where a key reason that arose for not wanting to pursue a science job was a perception of danger (which came up equally from boys and girls). Despite the prevalence of danger discussions in focus groups, this did not come up as a significant answer in the questionnaire data.

This would suggest that there is not a significant gender difference in the perceived danger of a science job – both girls and boys had concerns about the danger element. It should be noted that a possible reason that the same results were not observed in the questionnaire data could be because the question was phrased a little differently. In the focus group students were asked specifically to say 'What puts other people off science?' as opposed to the more open-ended question in the questionnaire; 'Why do you think that [you might not like a job in science]?'

**Similarity to a role model vs aspiration to their job**
By asking audiences which role model they felt had something in common with them, and then which role model had the job they would most enjoy, we wanted to see if we could evidence any pattern between the two as suggested in previous research (Gladstone & Cimpian, 2021). As our audiences only 'meet' the role models via facilitated videos we felt it would be interesting to check any effect from using 'indirect' role models.

| Which superhero has… | something in common with you? % | | job you would most like? % | |
|---|---|---|---|---|
| | Boys | Girls | Boys | Girls |
| Harriet (astronomer) | 9.9 | 20.9 | 9.7 | 30.9 |
| Lola (civil engineer) | 12.2 | 6.8 | 9.7 | 4 |
| Hannah (medical engineer) | 1.1 | 15.5 | 5.9 | 22.8 |
| Libby (environmental scientist) | 13.8 | 39.9 | 8.6 | 24.2 |
| Evan (electronics engineer) | 33.1 | 10.1 | 42.5 | 11.4 |
| Rory (flight engineer) | 29.8 | 6.8 | 23.7 | 6.7 |

*Green = most frequent, red = least frequent*
*Table 3: Heat map of % responses to which role model has something in common, and the job most liked (by gender)*

For the female students the environmental scientist (female) was most frequently stated as the person who had something in common with them, and the astronomer (female) had the job most female students said they would like. The civil engineer (female) was least frequently stated by girls as having anything in common with them and was also selected as the least desirable job (equal with the flight engineer).

It is possible that the most similar person for females was not the person who had the most desirable job because (despite mentioning a passion for saving the planet) her video mentioned that her job involved a bus powered by poo! It is likely that this affected the desirability of that job.

For the male students the electronics engineer (male) had the most in common with, and had the most desirable job, whilst the astronomer (female) had the least in common. The medical engineer (female) was selected as having the least desirable job.

When comparing responses to these two questions overall, we found that 197 (58.3%) of participants indicated that the superhero they had the most in common with was also the superhero with the job they would most like to have.

We then explored the gender of the superheroes that participants felt they had most in common with and who had the most desirable job. Looking at the gender of participants, 63% (n=114) of male participants felt they had most in common with a male superhero and 86% (n=123) of females felt they had most in common with a female superhero. Similarly, 66% (n=122) of males most wanted the job of a male superhero, and 82% (n=122) of females most wanted the job of a female superhero. There was a significant correlation that respondents tended to choose a role model of the same gender whether they were saying who they shared things in common with, or who had the most desirable job. This effect was larger in female students.

**Reasons students gave for choosing specific role-models**

We used the data from the post-show questionnaire to explore how altruism and enjoyment factors related to the choice of role model (Table 4). Many more girls than boys chose altruistic reasons to justify their choice of role model. For boys there was a large majority of reasons given related to enjoyment. The most commonly stated factor for girls on choosing a role model was also the perceived enjoyment of their job, but the proportion who mentioned altruistic similarities (to the person) was much higher than for boys, and was almost equivalent to the enjoyment when explaining why they would like the role model's job.

| | **What did you have in common [with the superhero]?** | | **Why did you like their job?** | |
|---|---|---|---|---|
| **Answers linked to…** | Boys (%) | Girls (%) | Boys (%) | Girls (%) |
| **Altruism** | 7 | 29 | 9 | 43 |
| **Enjoyment** | 64 | 46 | 66 | 45 |

*Table 4: Altruism and enjoyment by gender as reasons for choosing role models*

**Attitudes to scientists based on words chosen**

Before and after the show, questionnaires asked students to provide three words to describe scientists and engineers. These were categorised into positive, negative and neutral words using themes in NVivo and results are shown in Fig 3.. There was a clear increase in positive words used, and a reduction in negative words, from before the show to after the show. Both genders showed an increase in the number (and proportion) of positive words used after the show suggesting a shift in attitude about what scientists and engineers are like. There was a larger increase in the positive words from girls, but boys were more likely to give more 'extreme' views, where girls were more likely to be 'neutral'.

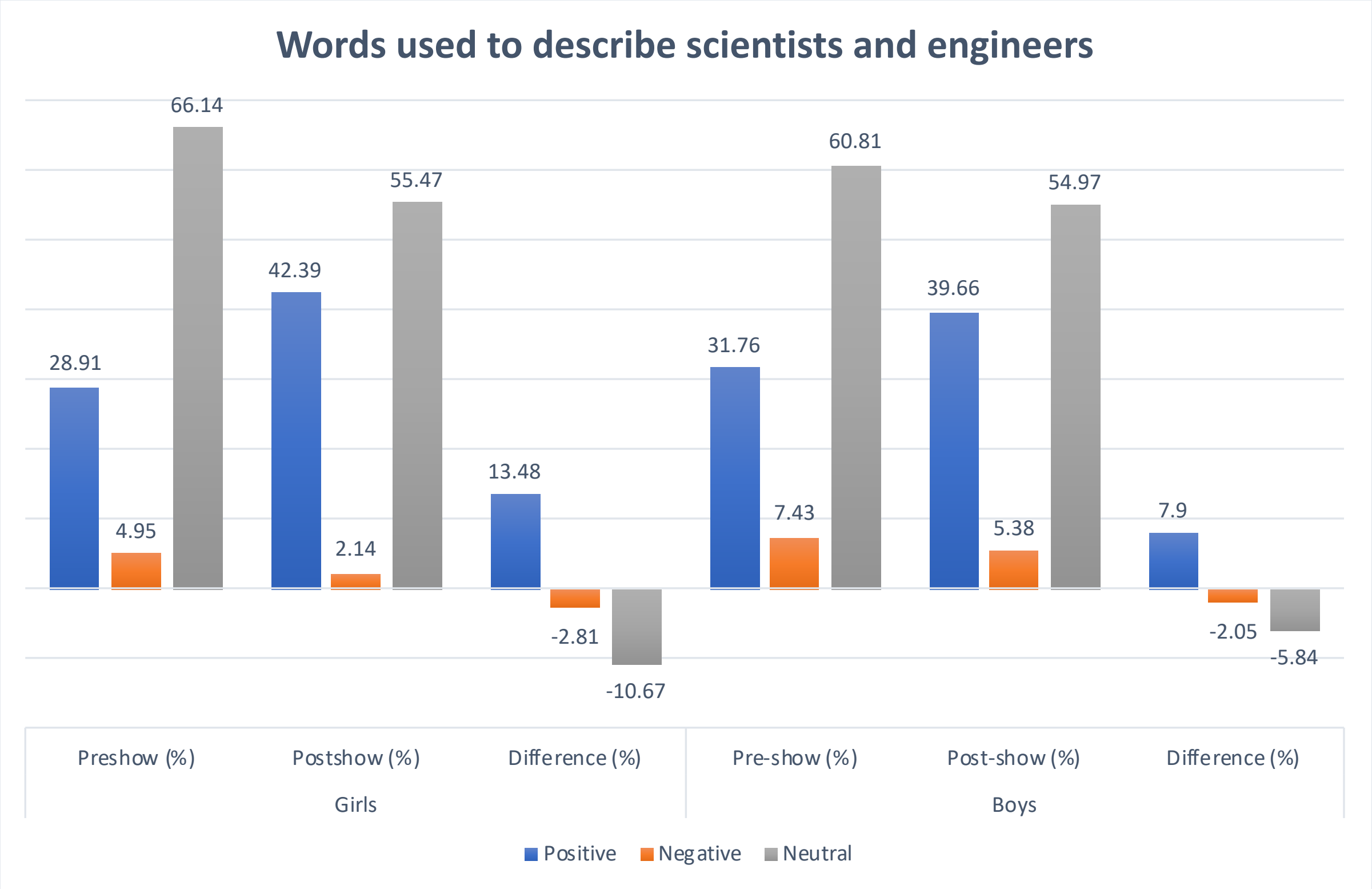

*Fig 3: Words used to describe scientists pre- and post-show (by gender)*

**Analysis of specific words used to describe scientists**

There was a decrease seen in the ‘crazy’, ‘nerdy’, and ‘evil’ stereotypes after the show. In the questionnaire response data from before the show roughly 5% of both boys and girls used these stereotypical words. After the show, there was a reduction in this perception for both boys and girls, but girls’ negative stereotypical attitudes to scientists decreased more than boys’ attitudes.

| | **Girls** | | | **Boys** | | |
|---|---|---|---|---|---|---|
| **Word groupings** | Pre-show (%) | Post-show (%) | Difference (%) | Pre-show (%) | Post-show (%) | Difference (%) |
| ‘Clever’ ‘Talented’ ‘Intelligent’ | 47.69 | 39.87 | -7.82 | 43.63 | 39.55 | -4.08 |
| ‘Awesome’ ‘Cool’ ‘inspiring’ | 10.12 | 21.51 | +11.39 | 11.49 | 18.36 | +6.87 |
| ‘Hardworking’ ‘Ambitious’ ‘Focussed’ | 6.30 | 5.91 | -0.39 | 6.56 | 7.10 | +0.54 |
| ‘Creative’ ‘Designers’ ‘Pioneers’ | 7.76 | 6.92 | -0.84 | 6.08 | 4.87 | -1.21 |
| ‘Dangerous’ ‘Brave’ ‘Strong’ | 4.39 | 3.65 | -0.74 | 5.89 | 3.75 | -2.14 |
| ‘Helpful’ ‘Important’ | 2.92 | 5.79 | +2.87 | 4.44 | 6.80 | +2.36 |
| ‘Nerdy’ ‘Crazy’ ‘Evil’ | 4.27 | 1.76 | -2.51 | 4.82 | 3.55 | -1.27 |

*Table 5 – Changes in specific descriptive words used before and after the show by gender*

Students’ awareness of both gender stereotypes and ‘nerdy’ stereotypes was apparent in the focus groups. In response to the question, ‘What puts people off a job in science?’ students observed that other people thought it was nerdy, even if they themselves did not hold this view.

**The ‘Clever’ stereotype**

The largest proportion of words used (both before and after the show) to describe scientists implied a certain type of person, someone ‘talented’ or ‘clever’, as shown in Table 5. Although there was some reduction in this stereotype, it was still the most common word listed. Girls originally had more of this view, but the reduction in the proportion of these words was greater for girls than for boys. It can be concluded that the show reduced the ‘clever’ perception of scientists for girls and boys. While boys’ perception of scientist’s cleverness reduced, they were influenced less than girls by the show.

**The altruistic view of STEM**

Words like ‘helpful’ and ‘important’ (and their synonyms) as a part of the three words to describe scientists were counted, as shown in Table 5. Although girls showed a larger increase in the ‘helpful’ perception, a higher proportion of boys’ responses reflected this view of scientists. Overall, the show did increase the ‘helpful’ perception of science, for both boys and girls, although there was a larger change in girls’ responses from a lower starting point.

**Conclusions**

The role model-based science show ‘Who wants to be a superhero?’ was shown to provide positive and diverse messages about the kinds of people who work in science and engineering. The use of role models in the show reduced the stereotypes about ‘nerdy’ and ‘crazy’ scientists, and perceptions that scientists and engineers are ‘clever’. This change was larger for girls’ than boys, though girls began with a higher proportion using these words in their descriptions. In the focus groups, it was clear that students were aware of both ‘nerdy’ and gender stereotypes of people working in STEM, yet their personal views tended not to be stereotypical. The show achieved a small increase in perceptions of the ‘helpful’ nature of science. However, the main reason for choosing a STEM career and identifying with a scientist, among both boys and girls was a shared interest or perceived enjoyment of the job.

As this case study used role models in an indirect way (through showing videos facilitated by a presenter rather than in person) it suggests that this could be an effective way to scale-up role model interventions in a sustainable and effective way.

A key finding from the focus groups showed that students may be put off a job in STEM because of its dangerous nature, as opposed to it being difficult or unattainable. Practitioners should be careful when balancing the choice of role model jobs shown in case what they think are ‘exciting’ jobs blur into jobs that can appear dangerous to students.

It is important that role models in any intervention talk about their personal interests and passions and explicitly contradict the negative stereotypes (such as scientists all being super-intelligent and nerdy). It is also important for role models to have a diversity of backgrounds and interests to make it more likely that each member of the audience will find someone that they resonate with. These findings reinforce that students need to see someone they relate to (whether by gender similarity or hobby/passion), and someone who shows that they enjoy their job to make it more likely that students may change their views on whether a job in science or engineering is for ‘people like them’.